# Novel indium phosphide charged particle detector characterization with a 120 GeV proton beam

Sungjoon Kim[1*], Manoj B. Jadhav[2], Vikas Berry[1], Jessica E. Metcalfe[2], and Anirudha V. Sumant[3]

1 Department of Chemical Engineering, University of Illinois at Chicago, 929 W Taylor Street, Chicago, Illinois 60608, USA
2 High Energy Physics Division, 9700 South Cass Avenue, Argonne National Laboratory, Lemont, Illinois 60439, USA
3 Center for Nanoscale Materials, 9700 South Cass Avenue, Argonne National Laboratory, Lemont, Illinois 60439, USA

*Corresponding Author E-mail: skim606@uic.edu
*Current affiliation: Applied Materials Division, 9700 South Cass Avenue, Argonne National Laboratory, Lemont, Illinois 60439, USA

**Abstract**

Thin film detectors which incorporate semiconductor materials other than silicon have the potential to build upon their unique material properties and offer advantages such as faster response times, operation at room temperature, and radiation hardness. To explore the possibility, promising candidate materials were selected, and particle tracking detectors were fabricated. An indium phosphide detector with a metal-insulator-metal (MIM) structure has been fabricated for particle tracking. The detector was tested using radioactive sources and a high energy proton beam at Fermi National Accelerator Laboratory. In addition to its simplistic design and fabrication process, the indium phosphide particle detector showed a very fast response time of hundreds of picoseconds for the 120 GeV protons, which are comparable to the ultra-fast silicon detectors. This fast-timing response is attributed to the high electron mobility of indium phosphide. Such material properties can be leveraged to build novel detectors with superlative performance.

**Introduction**

Recent advances in material science and deposition technology open a promising window for incorporating materials other than silicon and germanium into photon or charged particle detectors. Of the candidate materials, III-V compound semiconductors have the potential to perform as sensitive detectors that do not require cooling during operation. This translates to simpler device designs as well as operations since detector packaging and cooling systems are typically bulky, increase material that does not contribute to the detection, and necessitate additional support infrastructure[1]. Indium phosphide (InP) is a direct-bandgap compound semiconductor belonging to the group III-V materials, which has several properties that make it advantageous in particle tracking applications, such as high charge carrier mobility [2]. Due to its relatively large bandgap of 1.344 eV (at 300K), low number of free charge carriers ($n < 10^7 cm^{-3}$), and its high resistivity ($\rho > 10^8 \Omega\, cm$), room temperature operation of an intrinsic InP detector is theoretically possible [3]. However, intrinsic single crystals show n-type behavior with a high carrier concentration ($n \cong 10^{15} cm^{-3}$), which is thought to be the result of trace impurity donors such as Si and S [4]. To achieve sufficiently low conductivity, deep acceptors such as Cr, Co, and Fe are used to compensate for the residual donors in intrinsic InP[5, 6].

## Methods

### Device fabrication
Single crystal 2" 350 $\mu m$ thick single side-polished InP:Fe wafers manufactured at Pam-Xiamen were cleaved into 1 x 1 cm pieces and then cleaned in a piranha bath for 2 minutes at 120 °C. The front side of the chip was spin-coated with Microposit S1818 photoresist and then patterned using Heidelberg MLA150. A 4 x 4 array of 200 x 200 $\mu m$ pads were patterned on the front side of the chip. Electrodes were evaporated onto the chip using e-beam evaporation (Temescal FC2000). Chromium adhesion layer with a thickness of 10 nm was deposited, followed by an additional 100 nm of gold to allow wire-bonding. Backside electrodes were deposited using sputter coating (AJA Metal Sputtering System). A single backside electrode covers the entire chip, with a 20 nm and 100 nm thickness for chromium and gold, respectively.

### In-lab electrical and radiation source measurements
Current versus voltage (IV) measurements were conducted on the Form Factor Summit 11000 Probe Station. Keithley 4200A-SCS Parameter Analyzer was used to bias the device and collect electrical signals. Source measurements were conducted at room temperature using Sr-90, Tc-99, Ru-106, and Am-241 sources.

### Test beam measurements
A 120 GeV proton beam at Fermi National Accelerator Laboratory was used. The beam consisted of 350k counts per spill, with a spill duration of 4 seconds for every 1 minute. High-voltage power supply (CAEN, DT1471ET) was used to bias the device. Oscilloscope (Keysight, DSOS204A Infiniium S-Series) was used to collect the signal response.

## Results and Discussion

### Device design and fabrication
Three candidate materials were chosen for detector fabrication depending on the materials' potential as a particle tracker: iron-doped semi-insulating indium phosphide (InP:Fe), zinc-doped cadmium telluride (CdZnTe), and single crystal diamond. The candidate detector materials were selected due to their high carrier mobility and long carrier lifetime, as both contribute to the likelihood of generated carriers reaching the electrodes to form a measurable signal. CdZnTe and diamond have seen uses for particle detection and serve as a benchmark for comparing InP:Fe detector performance. The properties of the three detectors are summarized in table 1.

*Table 1. Device properties of charged particle detectors.*

| Material | Charge carrier type | Sensor layer thickness [$\mu m$] | Applied bias [V] | Front side electrode dimensions |
|---|---|---|---|---|
| InP | Electron | 350 | -250 | |
| $Cd_{0.96}Zn_{0.04}Te$ | Electron | 1000 | -650 | $200 \times 200\ \mu m$ |
| Diamond | Hole | 500 | 700 | |

For InP detectors, three different device structures are possible depending on the dopant type and concentration. First, a p-n junction can be formed using InP doped with zinc (p-type) and sulfur (n-type). The formation of an abrupt junction can be a challenge, and the thickness of the depletion region during reverse biasing will typically be smaller as the high concentration of dopants reduces the depletion region thickness. Second, a p-i-n setup may be used. This can significantly increase the thickness of the depletion region, and hence generate a stronger signal. The increase of the depletion layer thickness comes from the intrinsic layer sandwiched between the p and n type layers. Since the dopant concentration in the intrinsic layer is much smaller compared to the p and n type layers, the entire intrinsic layer may be depleted by reverse biasing. The final configuration is an iron-doped InP in a metal-insulator-metal (MIM) structure. This translates to the entire device bulk being depleted while also benefitting from simple device structure and fabrication process. Estimation of depletion region thickness is shown in figure 1. The depletion layer thickness was calculated using the following equation:

$$W = \left[\frac{2K_s\varepsilon_0}{q}\left(\frac{N_A + N_D}{N_A N_D}\right)(V_{bi} - V_A)\right]^{\frac{1}{2}}$$

Where
$W$: depletion width in $\mu m$
$K_s$: dielectric constant of InP
$\varepsilon_0$: vacuum permittivity
$q$: electron charge
$N_D$: n-type doping concentration
$N_A$: p-type doping concentration
$V_{bi}$: built-in potential
$V_A$: applied bias

Based on the calculations, the MIM structured detector was chosen and fabricated.

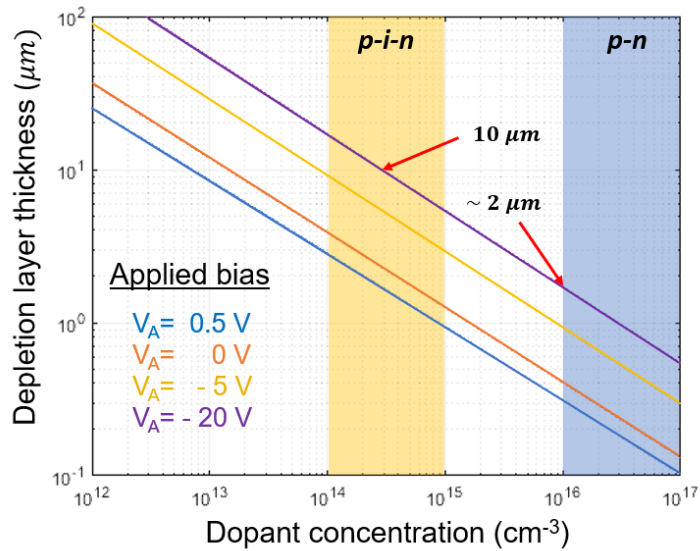

*Figure 1. Estimated depletion region thickness of InP devices with different junction configurations. X-axis represents the doping concentration of the lightly doped material. Yellow and blue regions represent typical concentrations of dopants.*

**In-lab electrical and radiation source measurements**
Figure 2 illustrates the testing setup. Following detector fabrication, the detectors are mounted on a readout board designed for ultrafast silicon detector (UFSD) testing. This single-channel readout board includes a low noise inverting amplifier with a wide bandwidth (~2 GHz). The inverting amplifier is connected to a second-stage 20 dB external amplifier (mini-circuit TB-409-52+) with a gain of 10. The read-out circuit's total trans-impedance was about 4700. Details on the readout board can be found in a previous work [7]. The detectors are HV-biased from the backside, and the front end collects the signal from the top pixels, which is further processed using a discrete amplifier.

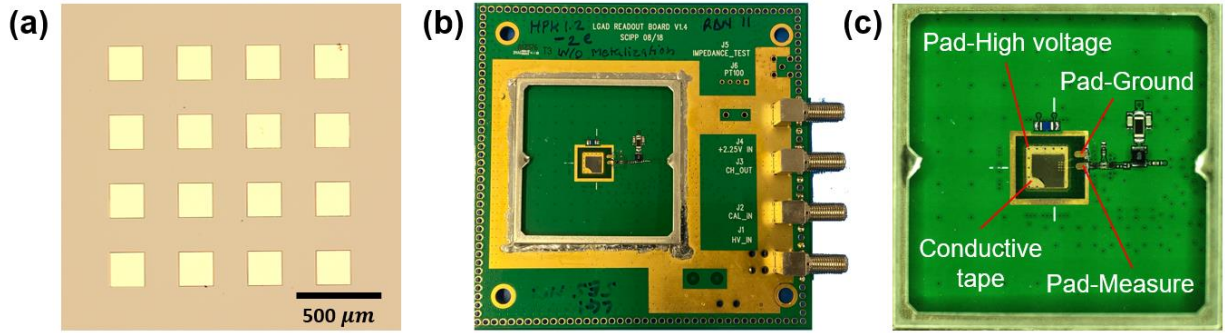

*Figure 2.*(a) Optical microscopy image of patterned electrodes on InP (front side) (b) wire bonded electrodes on InP detector (c) ASIC setup with built-in amplifier.

The InP:Fe device shows a typical MIM device response in an IV measurement (figure 3). Lower temperatures result in lower conductivities, confirming the insulating property of InP:Fe and indicate that cooling can reduce the thermal noise of the device. The IV measurements clearly show that the InP detector without a p-n junction does not need reverse biasing and can operated with either polarity. Slight variations between the negative and positive current are attributed to the difference in hole/electron mobility [8].

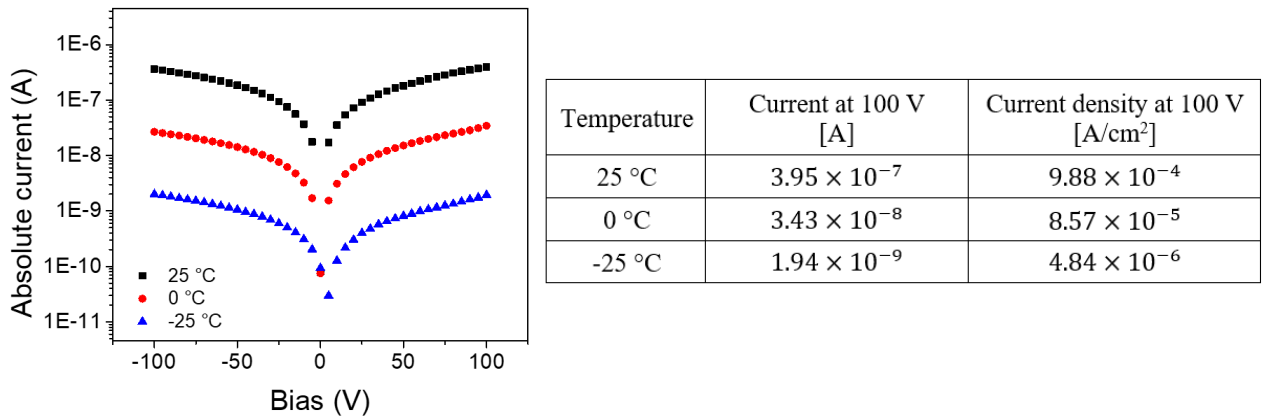

| Temperature | Current at 100 V [A] | Current density at 100 V [A/cm$^2$] |
|---|---|---|
| 25 °C | $3.95 \times 10^{-7}$ | $9.88 \times 10^{-4}$ |
| 0 °C | $3.43 \times 10^{-8}$ | $8.57 \times 10^{-5}$ |
| -25 °C | $1.94 \times 10^{-9}$ | $4.84 \times 10^{-6}$ |

*Figure 3.* I-V temperature study of InP:Fe device.

The current density calculated from the InP device shows comparable values to reported works [9, 10]. The differences between studies are attributed to the difference in the electrode sizes, dopant concentrations, and thicknesses of the devices.

Radiation source measurements were taken using different radioactive sources to estimate the energy resolution of the InP detector. The details of used radiation sources are summarized in table 2.

*Table 2.* *Activity and energy of radiation sources used in this study.*

| Radiation source | Activity [kBq] | Energy [keV] |
|---|---|---|
| Sr-90 | 370 | 546 |
| Rc-99 | 37 | 293.6 |
| Ru-106 | 40 | 621.9 |
| Am-241 | 37 | 59.6 |

The average collected charge was correlated to the energy of the incident particle, as shown in figure 4. The energy resolution is calculated using signal amplitude distribution fitted using Gaussian and Landau function convolution. The sigma value from the Gaussian fit provides full width at half maximum (FWHM) value as $2.355 \times sigma\ (\sigma)$, and Energy resolution is given by $\frac{FWHM \times 100}{Mean}$ (%). The energy resolution measured using fabricated InP detector is approximately 28%, as shown in figure 5. Figure 6 (left) shows that the signal amplitude increases with applied bias voltage, reflecting the signal-to-noise ratio (S/N) increase with a reasonably stable RMS noise. The signal amplitude rise is minimal compared to the wide range of operating voltage, indicating that the InP can operate at a lower bias without losing much of the signal quality.

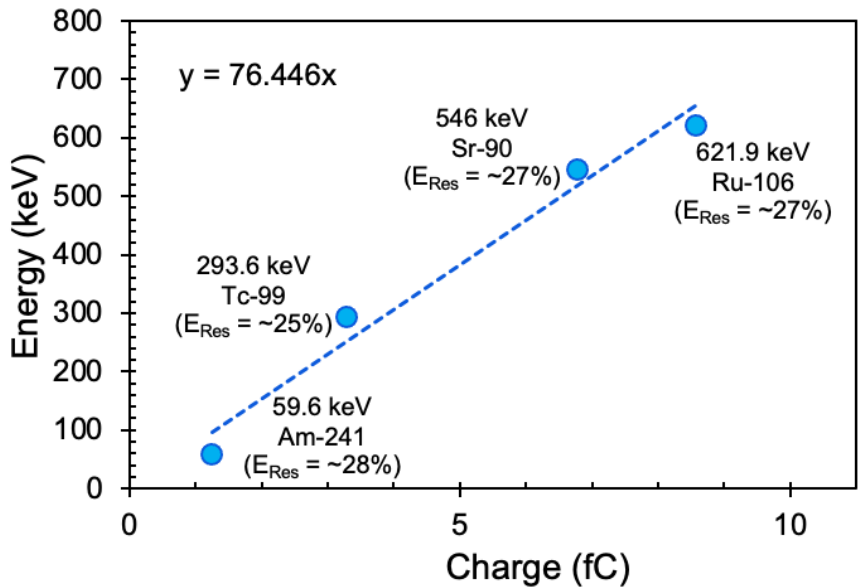

*Figure 4.* *Energy calibration of InP:Fe device using different energy sources at room temperature.*

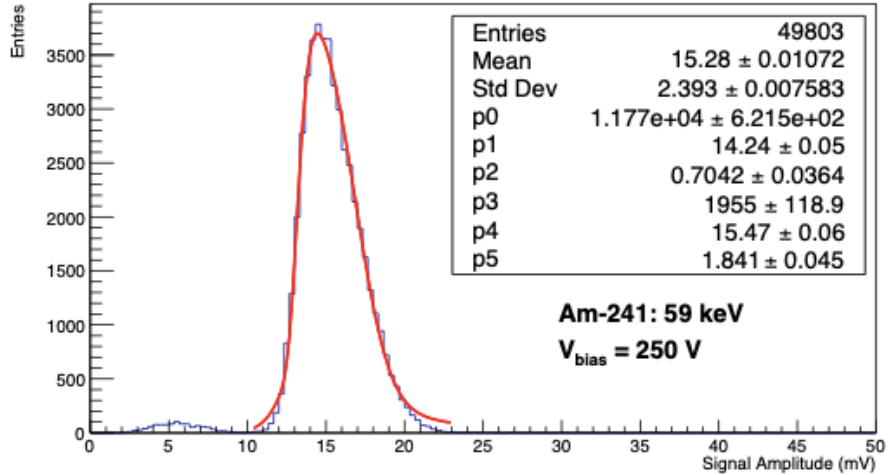

*Figure 5. Signal amplitude distribution for 59 keV peak from Am-241. The energy resolution is calculated to be ~28%.*

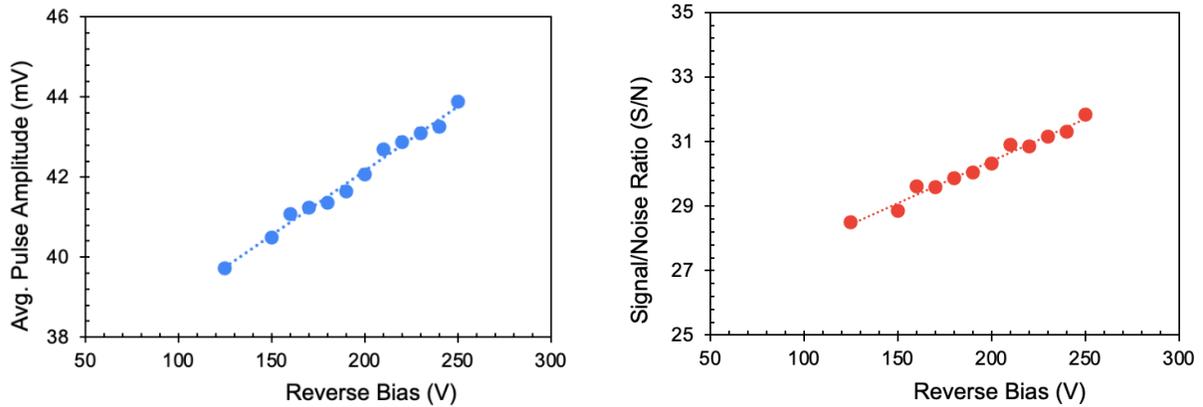

*Figure 6. Average pulse amplitude increases with bias voltage (left) which leads to an increase in Signal to Noise ratio (S/N) (right) (at 300 K).*

**Test beam measurements**

Test beam data is shown in Figure 7, where a bias sweep study was done on the InP detector. The bias was increased from 100 V to 250 V in 25 V increments. The detector was operated at room temperature and the leakage current was stable around 1.5 µA. The signal amplitude has very nominal variation over different bias voltages, which may be attributed to the insulating (or depleted) nature of InP and the fact that the test beam provides minimum ionizing particles, which deposit almost the same energy when traversing the detector. A similar trend is observed for the rise time, which is defined as the time required to reach 90% of maximum peak height from 10%. InP shows very fast response, with a rise time of around 250 ps, which is comparable to the ultra-fast silicon detectors which have been optimized for fast response time with additional gain layer in the bulk of silicon [7]. The signal amplitude shows a Landau distribution, which is typical for when charged particles lose energy through ionization in a thin layer of matter [11]. The Landau fluctuation in an ionizing particle detector originates from the knock-on electrons, which have gained sufficient energy from the interaction to become an ionizing particle itself. Rise time and pulse amplitude were compared with single crystal diamond and CdZnTe detectors in table 3. Due to the detectors having different bias voltages and device thicknesses means that direct comparison between the materials is difficult. However, they do provide a rough benchmark for the InP detectors as all the detectors were operated close to the breakdown voltage. Previous works that have explored CdZnTe detectors [12–14] and diamond detectors [15–19] report similar range of material properties to the ones observed in this work.

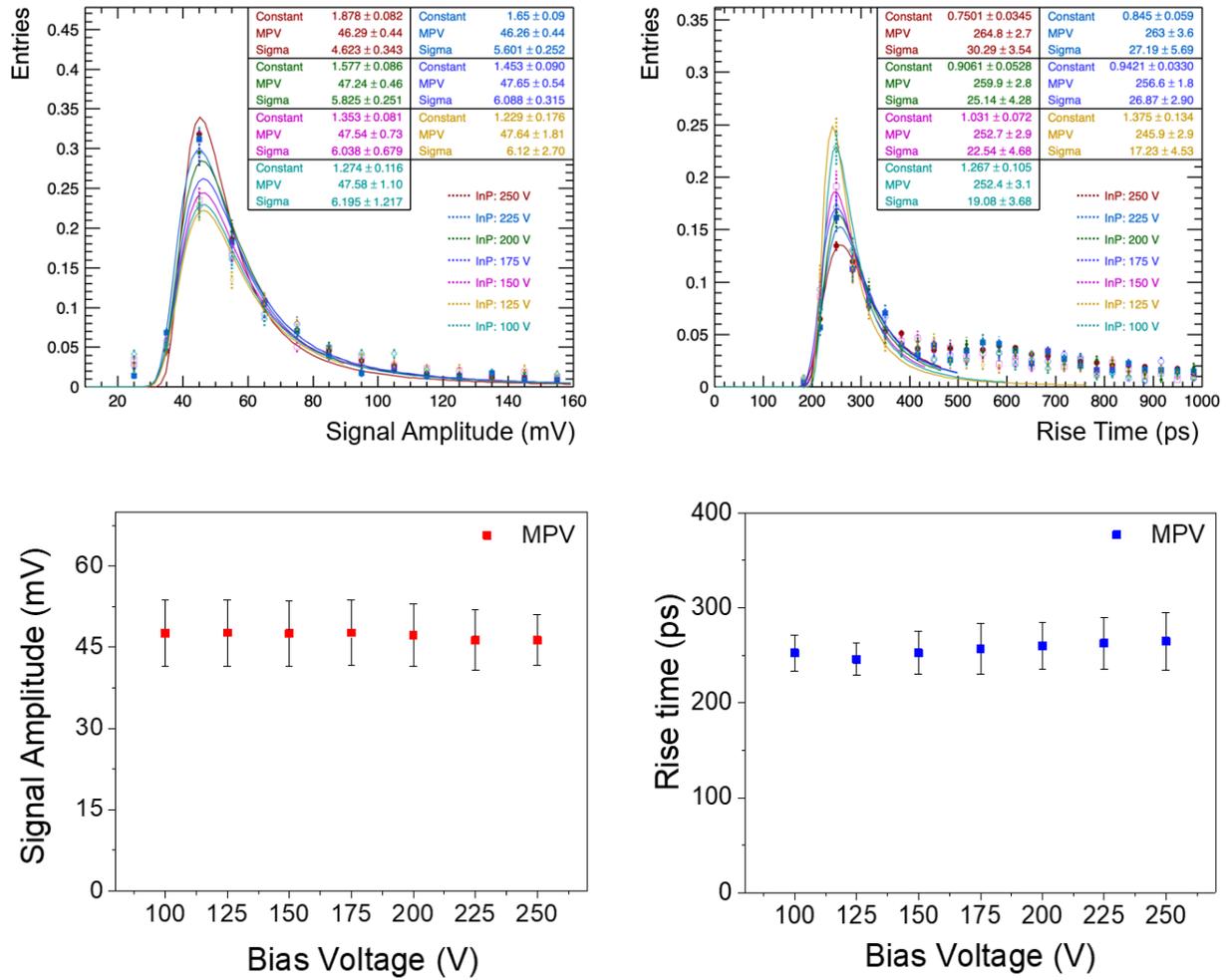

*Figure 7. Test beam data for InP detector.*

**Table 3.** Benchmark comparison with current state-of-the-art silicon detectors (at 300 K)

|  | **Crystalline Si** | **CdZnTe** | **Diamond** | **InP:Fe** |
|---|---|---|---|---|
| **Atomic number** | 14 | 43 | 6 | 32 |
| **Rise time (ns)** | - | 1800 | 300 | 0.25 |
| **Pulse amplitude (mV)** | - | 130 | 80 | 50 |
| **Carrier drift mobility (cm²/Vs)** | 1450 (e⁻)<br>500 (h⁺) | 1050 (e⁻)<br>90 (h⁺) | 1800 (e⁻)<br>2300 (h⁺) | 4600 (e⁻)<br>150 (h⁺) |
| **Carrier lifetime** | > 100 $\mu s$ (e⁻)<br>> 100 $\mu s$ (h⁺) | 0.1-2 $\mu s$ (e⁻)<br>0.1-1 $\mu s$ (h⁺) | ~ 100 ns (e⁻)<br>~ 50 ns (h⁺) | ~ 1 ns (e⁻)<br>~ 1 ns (h⁺) |
| **Key features** | Mature technology, long radiation length | High Z: suitable for X-ray detection, slow rise time | Long radiation length and high radiation tolerance | Very fast response, thick sensing layer possible through doping |

## Conclusion

Charged particle detectors were fabricated using semi-insulating crystalline indium phosphide wafers. The indium phosphide detector was benchmarked against other detectors that used CdZnTe and single crystal diamond. The detectors were tested using in-lab radiation sources as well as the 120 GeV proton test beam at the Fermi National Accelerator Laboratory. Indium phosphide detectors show great promise as charged particle detectors, showing very fast timing responses in the picosecond region. The spectroscopy shows linear energy calibration using 4 different energy sources with energy resolution of ~28%. The detector provides good performance for the first iteration with a nominal optimized design and architecture. The measured rise time is comparable to the fast silicon detectors such as low-gain avalanche diode (LGAD) detectors. Simplistic device design can translate to lower fabrication costs compared to some silicon detectors. Further improvements can be made by integrating indium phosphide with pre-existing CMOS technology, such as even faster response times and cleaner signals. The sensing layer does not need to be thinned down to very low thicknesses that can potentially cause durability issues, which is required for some silicon detectors for fast response.

**Acknowledgements**

Funding from Argonne National Laboratory, provided by the Director, Office of Science, of the U.S. Department of Energy under Contract No. DE-AC02-06CH11357. This work was supported by the Office of Naval Research (Contract: N000141812583) and University of Illinois at Chicago. Work performed at the Center for Nanoscale Materials, U.S. Department of Energy Office of Science User Facilities, was supported by the U.S. DOE, Office of Basic Energy Sciences, under Contract No. DE-AC02-06CH11357. This document is prepared using the resources of the Fermi National Accelerator Laboratory (Fermilab), a U.S. Department of Energy, Office of Science, Fermi Lab Test Beam Facility. Fermilab is managed by Fermi Research Alliance, LLC (FRA), acting under Contract No. DEAC02-07CH11359.